\documentclass{article}
\usepackage{latexsym}
\usepackage{sss}

\usepackage[pdftex]{graphicx} 
\usepackage{comment}
\usepackage{changepage}
\usepackage{amsmath,amssymb}
\usepackage[numbers]{natbib}
\begin{document}
\date{}
\title{\LARGE{\bf
Statistical Properties of Gray Color Categorization:\\ Asymptotics of Psychometric Function
}
}
\author{Ihor Lubashevsky, Marie Watanabe \\
University of Aizu, Tsuruga, Ikki-machi, Aizu-Wakamatsu City, Fukushima 965-8560, Japan\\
\\ 
E-mails: (IL) i-lubash@u-aizu.ac.jp,  (MW) s1190150@gmail.com	
}

\maketitle
\thispagestyle{empty}
\abstract{
The results of our experiments on categorical perception of different shades of gray are reported. A special color generator was created for conducting the experiments on categorizing a random sequence of colors into two classes, light-gray and dark-gray.	 The collected data are analyzed based on constructing (\textit{i}) the asymptotics of the corresponding psychometric functions and (\textit{ii}) the mean decision time in categorizing a given shade of gray depending on the shade brightness (shade number). Conclusions about plausible mechanisms governing categorical perception, at least for the analyzed system, are drawn.
}

\section{Introduction}

Humans cannot distinguish similar stimuli from one another or discriminate between them using some criterion if a difference in their intensity or other quantitative characteristics does not exceed a certain \textit{fuzzy} threshold. It makes human response or decision-making intrinsically discrete and indeterministic.

When a perceived stimulus is of physical nature and the goal is recognizing variations in its intensity, for example, intensity of sound or brightness, this threshold is mainly determined by human physiology. The corresponding phenomena and their description are the subject of psychophysics (see, e.g., \cite{gescheider2013psychophysics}). However, if a stimulus comes from the social domain, e.g., when we want to buy something and estimate its price in terms of ``cheap,'' ``acceptable,'' and ``expensive,'' human cognition contributes substantially to this process. It is also the case  when a stimulus is of physical nature but we need to classify it into a few categories to make some decision. This classification process is often called \textit{categorical perception}.  Categorical perception is a general term describing situations when  
%
\begin{adjustwidth*}{0.05\columnwidth}{0.05\columnwidth}
	we tend to perceive our world in terms of the categories that we have formed. Our perceptions are warped such that differences between objects that belong to different categories are accentuated, and differences between objects that fall into the same category are deemphasized \cite[p.~69, left column]{goldstone2010categorical}. 
\end{adjustwidth*}
%
The first roots of categorical perception research may be addressed to investigations at Haskins Laboratories after the construction of the first research-oriented speech synthesizer, the Pattern Playback \cite{harnad2005cognize}. Liberman, Harris, Hoffman, and Griffith  \cite{liberman1957discrimination} used this tool to construct a series of syllables spanning the three categories, \textbf{b}, \textbf{d}, and \textbf{g}, preceding a vowel approximating \textbf{e}. Although these stimuli formed a physical continuum obtained by increasing the onset frequency, subjects classified them into these three categories. 
For a review of the basics concepts and notions used in studying various phenomena of categorical perception a reader may be addressed to \cite{harnad2005cognize}.

Our research concerns a similar problem, categorization of colors. There is a vast amount of literature about various aspects of color categorization including fMRI investigations of brain activity. A comprehensive discussion of this problem and related ones  in cognitive science can be found, e.g., in \cite{cohen2005handbook}. 

In our previous experiments, described in short in the supplementary Section~\ref{Kobayashi}, shape recognition near perception threshold was examined based on the analysis of the \textit{asymptotics} of the corresponding psychometric function \cite{Kobayashi2014}.  Exactly this asymptotic analysis has enabled us to discriminate directly plausible mechanisms governing human recognition near perception thresholds and put forward a hypothesis about its \textit{emergent} nature. 
The purpose of present work was to verify whether the found features are more general, in particular, whether categorical perception, involving mental processes in addition to pure physiological ones, also exhibits similar properties. By  way of example, categorization of different shades of gray color was selected for investigation. In choosing this problem we have assumed that, on one hand, mental phenomena should play an essential role in this process and, on the other hand, the discrimination within one color according to its shades may be of simpler mechanism than that of different colors.

\section{Color generator}

\begin{figure*}[t]
	\begin{center}
		\includegraphics[width=0.8\columnwidth]{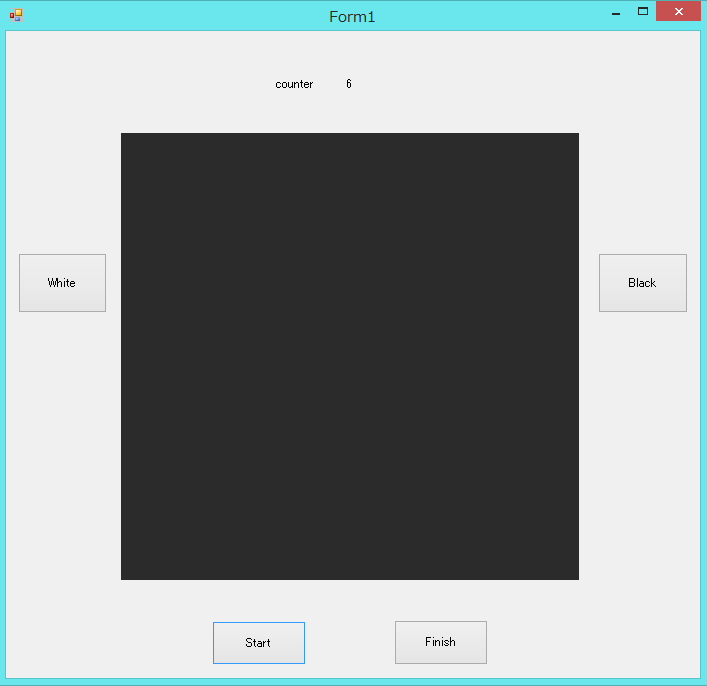}\hspace{1.5cm}
		\includegraphics[width=0.8\columnwidth]{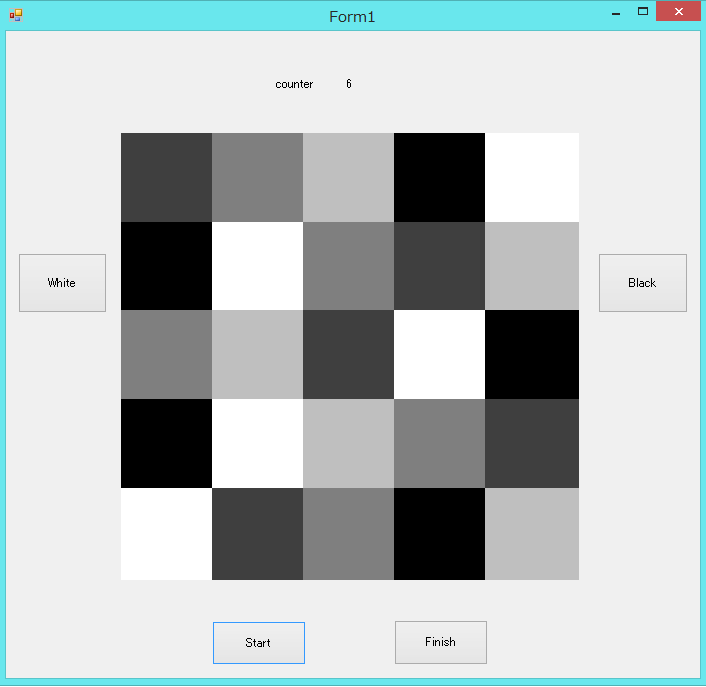}
	\end{center}
	\caption{Screenshot of the visualized window with the changeable color square. Left fragment is a color presented for categorization, the right fragment shows the color pattern between successive trials of color categorization, it is used for depressing possible effects of human iconic memory.}
	\label{Fig:CSim}
\end{figure*}

The following color generator software was developed.  On Lenovo LI2221s Monitor ($47.7\times26.8$~cm screen) a computer under the operating system Windows~8.1  visualizes  a window of size of $17\times16$~cm with a square $\mathbb{S}$ of size of $11\times11$~cm placed at its center. Color inside this square is changed during experiments; the remaining window part is filled with a neutral gray, namely, RGB(240,240,240). The brightness and contrast of the screen was set to 70\% and 60\%, respectively, using the monitor on-screen display controls, the other monitor setup characteristics were default. The gamma correction was implemented using the Windows built-in calibration tool manually based on subject's visual perception, the other color characteristics (brightness, contrast, and color balance) were not modified via the built-in calibration tools. 

Human-computer communication was implemented via a standard game joystick (USB joystick Elecom JC-U2912). Each trial of color categorization is implemented as follows. A random integer $I\in[0,255]$ is generated and the area $\mathbb{S}$ is filled with the gray color $G(I) := \mathrm{RGB}(I,I,I)$. Then a subject has to classify the visualized gray color  $G(I)$ according to his/her perception into two possible categories, ``light gray'' (LG) and  ``dark gray'' (DG). A made choice is recorded via pressing one of two joystick buttons. Then a mosaic pattern of various shades of gray shown in Fig.~\ref{Fig:CSim} (right fragment) is visualized for 500~ms. This mosaic pattern is used to depress a possible interference between color perception in successive trials that can be caused by human iconic memory (visual sensory memory) or, speaking more strictly, by the visible persistence component of iconic memory; its duration can be evaluated as 100-200~ms \cite{long1980iconic}.  After that a new number $I$ is generated and the next trial starts. 
The generated number $I_k$, the made choice $B_k$ (ID of the pressed button), and the time interval $\delta_k$ between the moment when the color $G(I_k)$ was visualized and the moment when a subject pressed the button are recorded as the data values of trial $k$.    

To analyze asymptotics of the psychometric identification/discrimination functions, a large amount of statistical data is required. So the generation of random integers $I$ is optimized dynamically in run-time as it is described in Appendix~\ref{App:B}. The programming of the color generator operations was made under C\# language using Microsoft Visual Studio Ultimate 2012.

\section{Experiment setup}

In order to control the environment conditions during experiments spanning over several days, the blinds of windows near the used monitor were closed and all fluorescent lights were turned on. The same computer was used for all the experiments. Participants were instructed to sit in front of the monitor at distance about 70~cm in relaxed posture. Two joystick buttons were used for fixing a choice made in categorizing a given color, the left one matched the LG choice, the right one matched the DG choice. The visible labels ``White'' and ``Black'' on the window were arranged in the same order (Fig.~\ref{Fig:CSim}) to prompt a subject the appropriate joystick button for pressing.  In instructing the subjects the terms \textit{light gray} and \textit{dark gray} instead of \textit{white} and \textit{black} were used to name the two color categories of the required color classification. More than half a century ago, Benjamin Lee Whorf \cite{whorf1956language} proposed that language affects perception and thought and is used to segment nature, a hypothesis that has since been tested by linguistic and behavioral studies. Recently based also on functional magnetic resonance imaging technique, it has been demonstrated that easy-to-name and hard-to-name colors influence the brain activity of subjects performing perceptual decision during color discrimination tasks \cite{Tan11032008,TingSiok19052009}. For this reason the color names more relevant to visualized colors were chosen.

The experiments were set up as follows. Four subjects, two female and two male students of age 21--22 were involved in these experiments. The experiments spanned 5 successive days, for each subject the total number of data records planed to be collected is 2\,000 per day and finally 10\,000 for 5 days. One day set comprised four blocks of 15~min experiments separated by 3~min rest, if 2000 records were collected earlier on this day experiments were stopped.  For each subject the total data set file aggregates all the records collected during 5 days.     

\section{Categorization mechanisms}

There is a vast literature devoted to psychometric functions and their mathematical models (see, e.g., \cite{wichmann2001psychometric1,wichmann2001psychometric2,zchaluk2009model,prins2012psychometric}). In the present research, however, the main attention is focuses on the asymptotic behavior of the psychometric function because only it can reflect the details of the human recognition mechanisms. So let us consider two plausible physical mechanisms of decision-making near perception threshold giving rise to different asymptotic behavior.

\subsection*{Mechanism 1: Noise signal detection}

This mechanism can be represented in term of noise signal detection. In a simplified form it implies the following. A physical stimulus of intensity $\epsilon$ is converted by the human brain into some ``internal'' stimulus with a certain noise. Because we deal with relatively small variations in the stimulus intensity near the corresponding persecution threshold  $\epsilon_\text{thr}$ let us ignore the difference between the physical and perceived stimuli. In the case under consideration this difference is responsible only for some renormalization of unknown constants. So we may assume that a subject perception of the given stimulus is described by the expression
\begin{equation}\label{int:1}
\epsilon_h = \epsilon + \Delta \xi\,,
\end{equation} 
where $\xi$ is a white noise with unit intensity, i.e., a random variable with the Gaussian distribution function
\begin{equation*}
p_G(\xi) = \frac{1}{\sqrt{2\pi}}\exp\left\{-\frac{\xi^2}{2}\right\}\,,
\end{equation*} 
and $\Delta$ is the noise amplitude. This model claims a subject to perceive the stimulus when its intensity $\epsilon_h$ exceeds some the threshold $\epsilon_\text{th}$, i.e.,
\begin{equation}\label{int:2}
\epsilon + \Delta \xi > \epsilon_\text{th}\,.
\end{equation} 
Then the probability $P(\epsilon)$ of a subject recognizing this stimulus is
\begin{equation}\label{int:3}
P(\epsilon) = \int_{-\infty}^{(\epsilon_\text{th} - \epsilon)/\Delta}p_G(\xi)d\xi = \frac12\text{erfc\,}\left(\frac{\epsilon-\epsilon_\text{th}}{\sqrt2\Delta}\right)\,.
\end{equation} 
Here $\text{erfc\,}(x)$ is the complementary error function specified by the expression
\begin{equation*}
\text{erfc\,}(x) = \frac2{\sqrt\pi}\int_x^\infty e^{-x^2}dx
\end{equation*} 
with the asymptotics  
\begin{equation}\label{int:6}
\text{erfc\,}(x) \sim
\begin{cases}
\dfrac{1}{x\sqrt{\pi}}e^{-x^2},& \text{if $x\to\infty$,}\\[0.5em]
2-\dfrac{1}{|x|\sqrt{\pi}}e^{-x^2},& \text{if $x\to-\infty$.}
\end{cases}
\end{equation} 
Exactly these expressions describe the asymptotics of the psychometric function $P(\epsilon)$ at the beginning and the end of the transition region wherein the function $P(\epsilon)$ exhibits the main change from 0 to 1. 

\subsection*{Mechanism 2: Mental stochastic process}

As a plausible alternative to the first mechanism, let us note the concept of noise-driven switching widely used for describing bistable cognitive phenomena, e.g., perceptual categorization \cite{tuller1994nonlinear,van2002non}, decision making under risk \cite{van2013modeling}, as well as the noise-activation of human intermittent control in pendulum balancing \cite{zgonnikov2015double}. The cognitive process implementing categorization perception as decision-making in choosing between two options is described by some random variable $\zeta(t)$ whose stochastic dynamics is governed by a certain double-well potential $U(\zeta,\epsilon)$. Without lost of generality we may assume that the two values $\zeta = 0$ and $\zeta = 1$ correspond to the bottoms $U_0(\epsilon)$ and $U_1(\epsilon)$ of the wells and the probability of finding the system at one of the wells gives the probability of choosing the corresponding option. First, this stochastic mental processes in the subject's mind is activated by the goal of decision-making with respect to the expected stimulus. Second, the physical stimulus affects the cognitive process endowing the potential $U(\zeta,\epsilon)$ with dependence on the stimulus intensity $\epsilon$ as a parameter. So the probability of choosing option $i$ ($i=0,1$) can be written as 
\begin{equation}\label{mec2:1}
P_i(\epsilon) = \frac{\exp\{-U_i(\epsilon)\}}{\exp\{-U_0(\epsilon)\}+\exp\{-U_1(\epsilon)\}}\,.
\end{equation}  
It is quite natural to define the perception threshold $\epsilon_\text{th}$ via the equality $U_0(\epsilon) = U_1(\epsilon)$ for $\epsilon =\epsilon_\text{th}$. Then expanding the functions $U_0(\epsilon)$, $U_1(\epsilon)$ of the argument $\epsilon$ in the Taylor series and confining the expansion to the leading terms we can rewrite \eqref{mec2:1}, for example, for $i=0$ as 
\begin{equation}\label{mec2:2}
P_0(\epsilon) = \frac12 \left\{1+\tanh\big[\beta(\epsilon - \epsilon_\text{th}) \big]  \right\} \,,
\end{equation}  
where $\beta$ is some constant. 

It should be noted that exactly this mechanism can be regarded as emergent one because it assumes the existence of a common mental process of decision-making affected by a specific stimulus via the dependence of the double-well potential on the stimulus intensity.

\section{Data analysis}\label{sec:DA}

The collected data have been used in constructing three functions. The first one is the psychometric function for the light-gray category, i.e., the probability $P_w(I)$ of choosing the light-gray category for the shade of gray, $G(I)$, specified by a given integer $I$. The second one, $P_b(I)$, is actually the same function for the dark-gray category. Because in these experiments the choice should be made between only two categories, for each value of $I$ the equality 
\[
P_w(I) + P_b(I) = 1
\]
holds. The third function, $T(I)$, is the mean duration time of making decision during one trial of categorization  for a given shade $G(I)$ of gray.      

Based on the collected data, e.g., the psychometric function for the LG category is constructed according to the following formula
\[
P_w(I) = \left[ \sum_{k=1}^M  \delta(I,I_k) \delta(0,J_k) \right]\cdot \left[ \sum_{k=1}^M  \delta(I,I_k) \right]^{-1},
\]
where $k$ is the index of record line in the data file which runs from 1 up to $M=10\,000$ (total number of records for 5 days) and $\delta(i,j)$ is the Kronecker delta:
\[ 
\delta (i,j)
=
\begin{cases}
1, & i=j\,,\\
0, & i\neq j\,.
\end{cases}
\]
The function $T(I)$ is specified via the expression
\[
T(I) = \left[ \sum_{k=1}^M  \delta(I,I_k) \Delta_k \right]\cdot \left[ \sum_{k=1}^M  \delta(I,I_k) \right]^{-1}.
\]
These functions were constructed for each subject. 

\newcommand{\figscale}{0.93}
\begin{figure*}
	\begin{center}
		\includegraphics[width=\figscale\columnwidth]{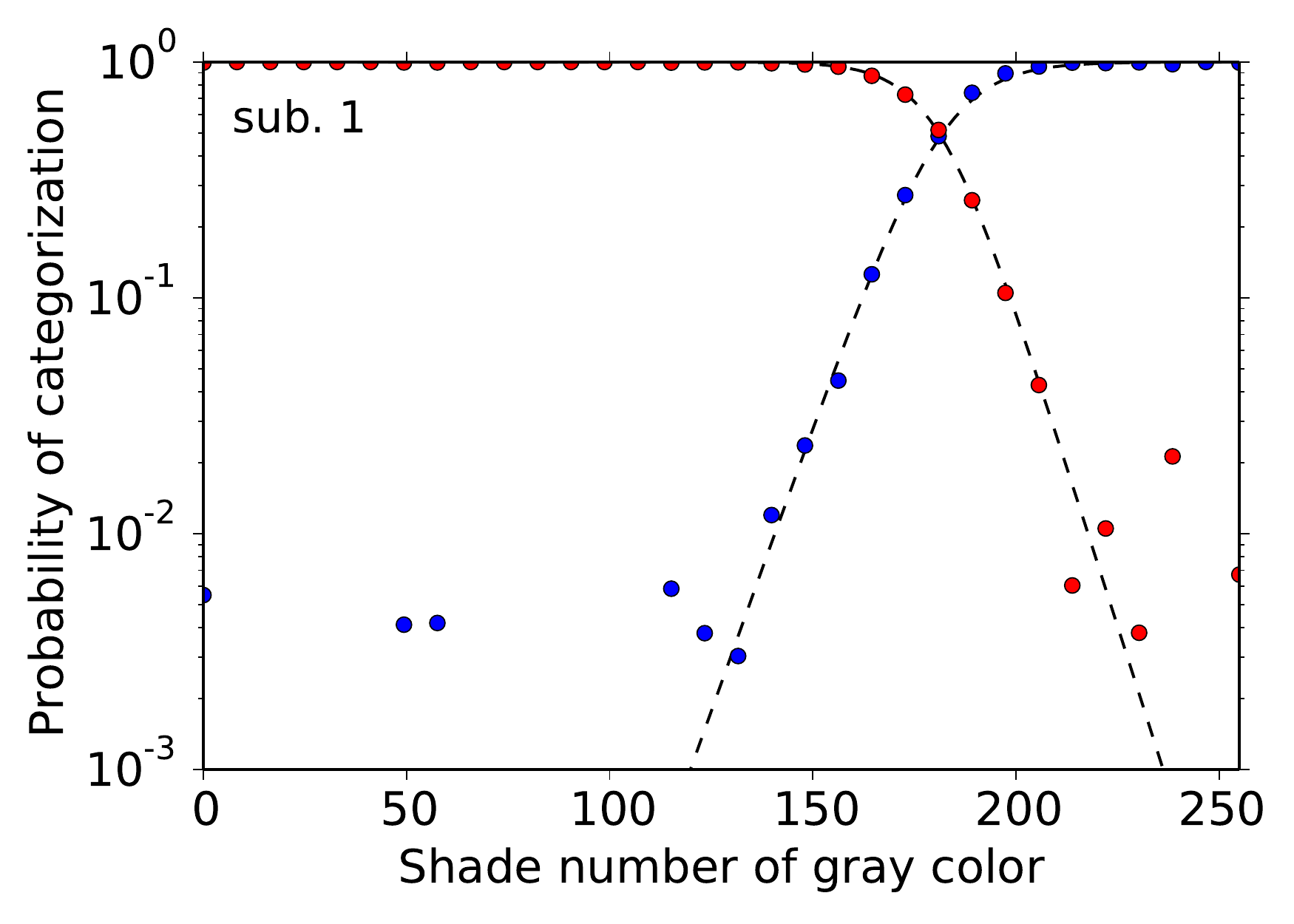}\quad
		\includegraphics[width=\figscale\columnwidth]{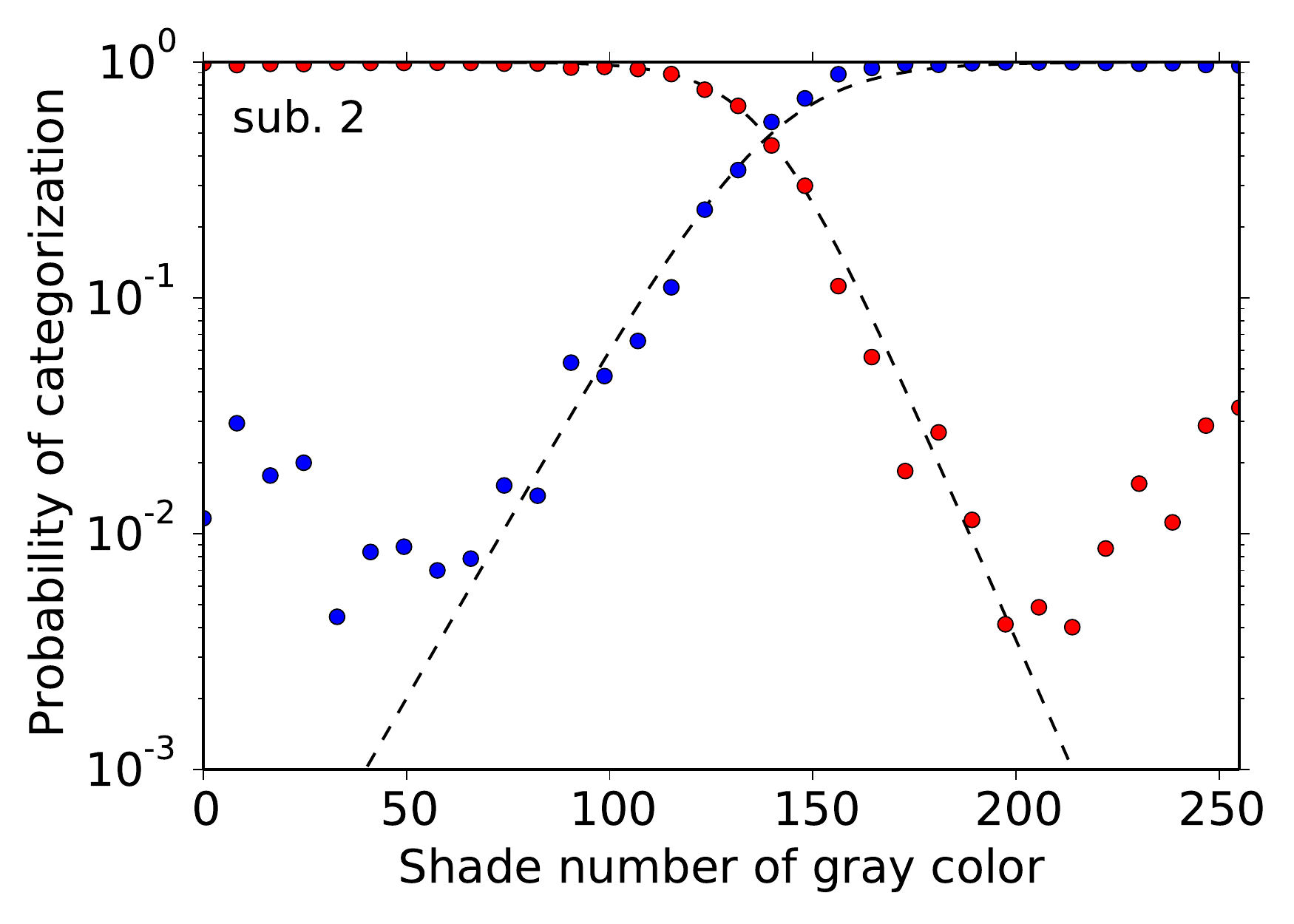}\\[-2.2mm]
		\includegraphics[width=\figscale\columnwidth]{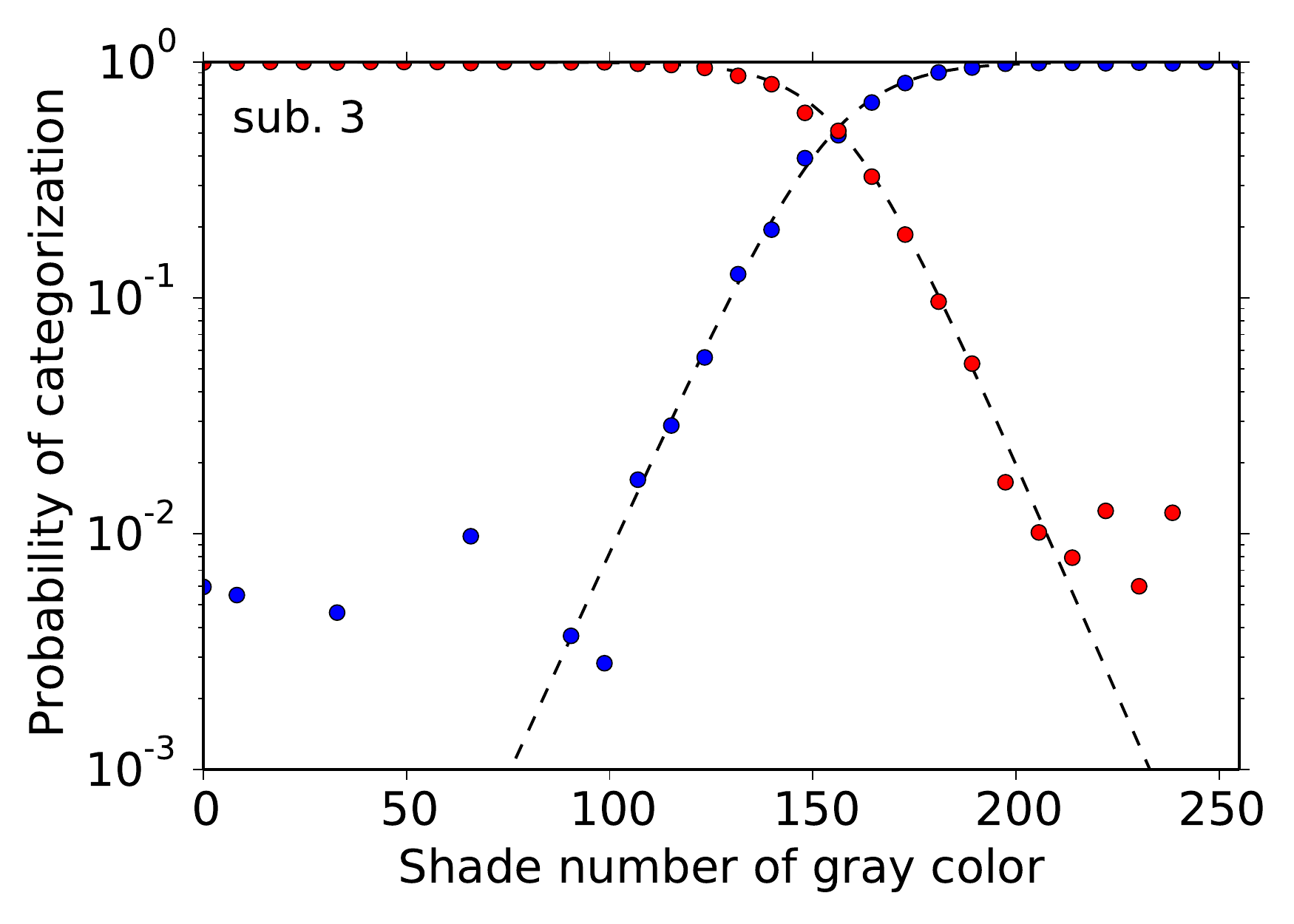}\quad
		\includegraphics[width=\figscale\columnwidth]{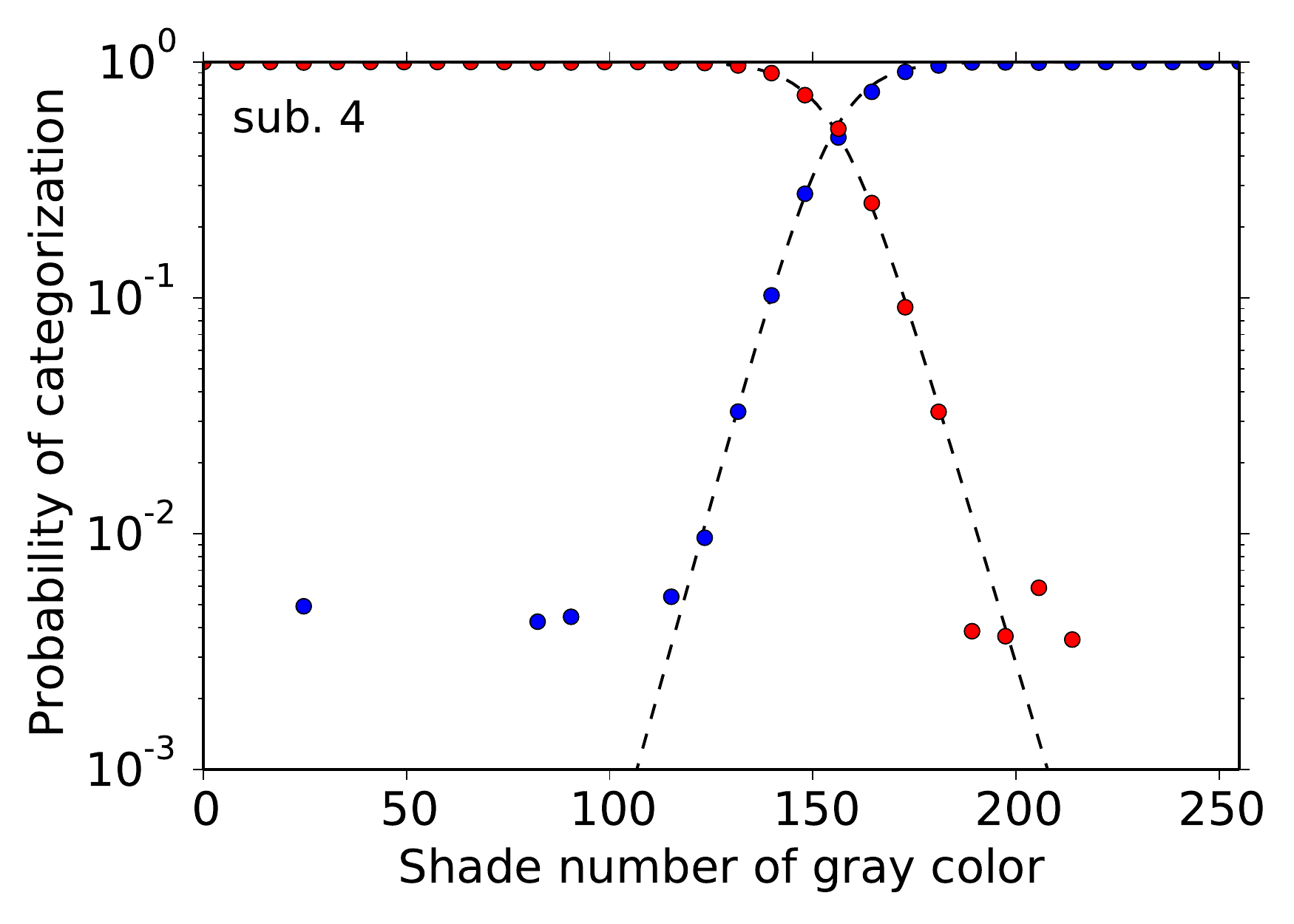}
	\end{center}
	\vspace*{-1.5\baselineskip}
	\caption{Psychometric functions of categorization of shades of gray, data points corresponding to the light-gray class and dark-gray class are shown in red and blue, respectively. Dashed lines represent fitting functions given by expression~(\ref{fitfun}) with parameters presented in Tab.~\ref{Tab:1}}
	\label{F3}
	\vspace*{-0.35\baselineskip}
	%
	\begin{center}
		\includegraphics[width=\figscale\columnwidth]{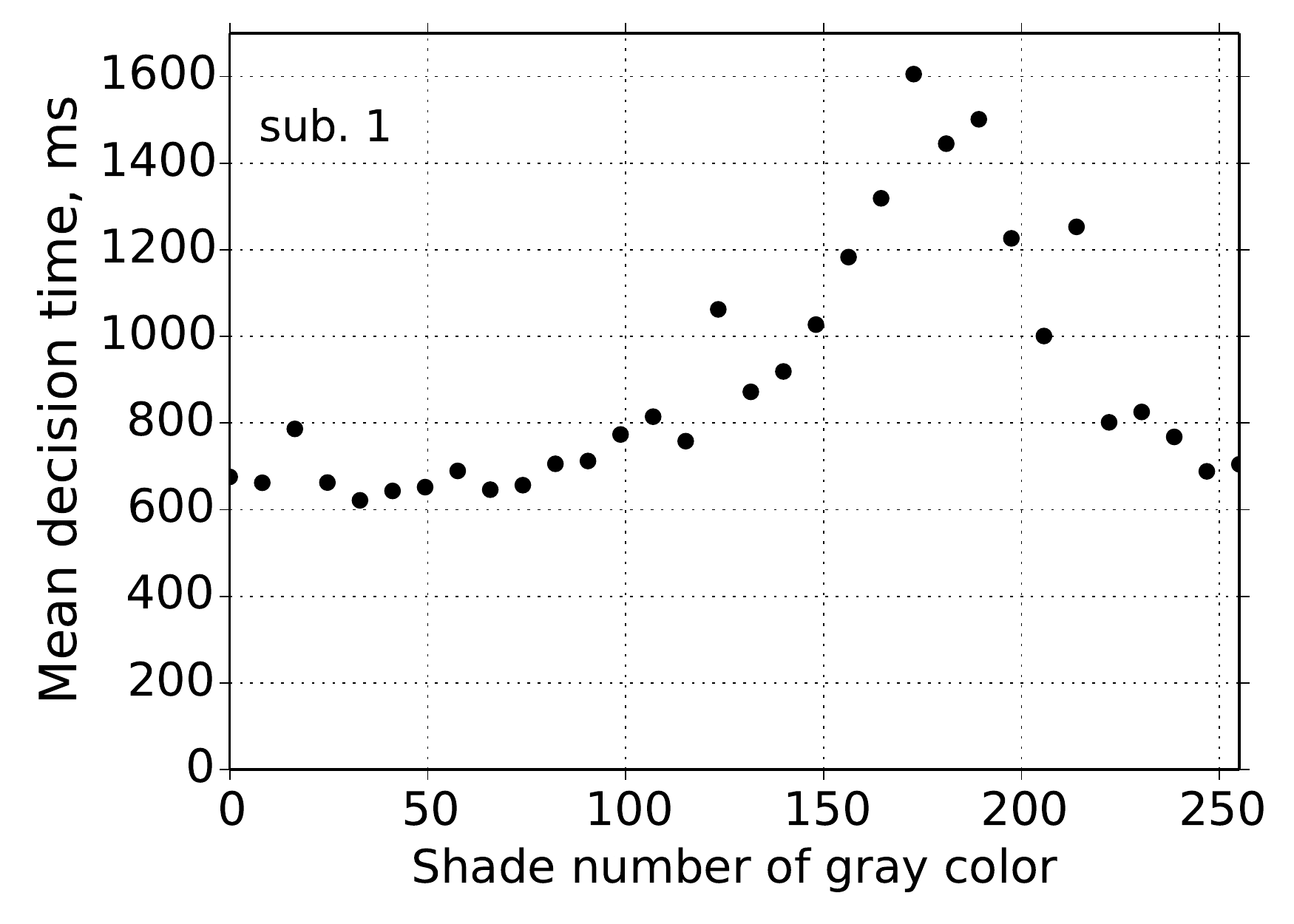}\quad\quad
		\includegraphics[width=\figscale\columnwidth]{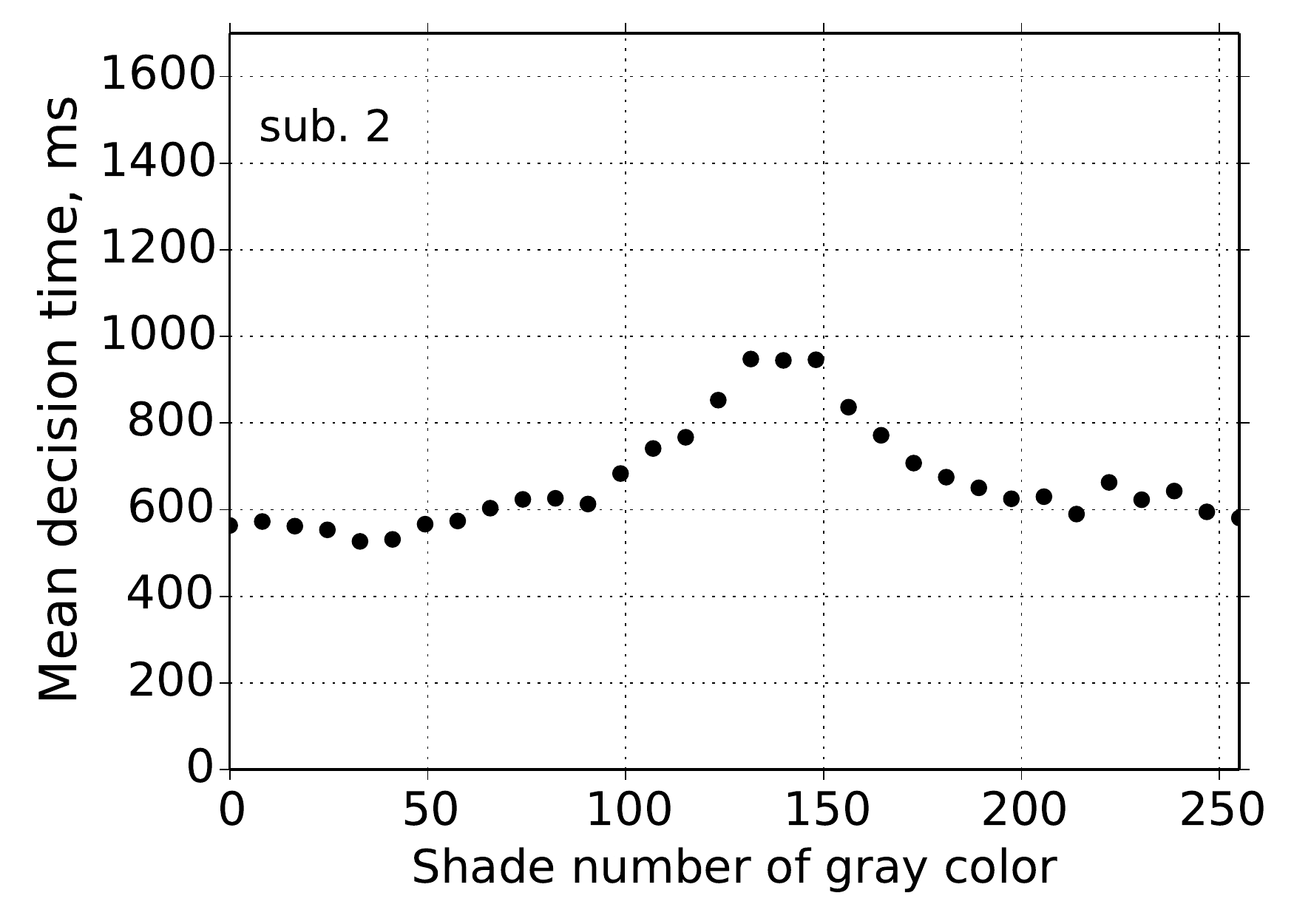}\\[-2.2mm]
		\includegraphics[width=\figscale\columnwidth]{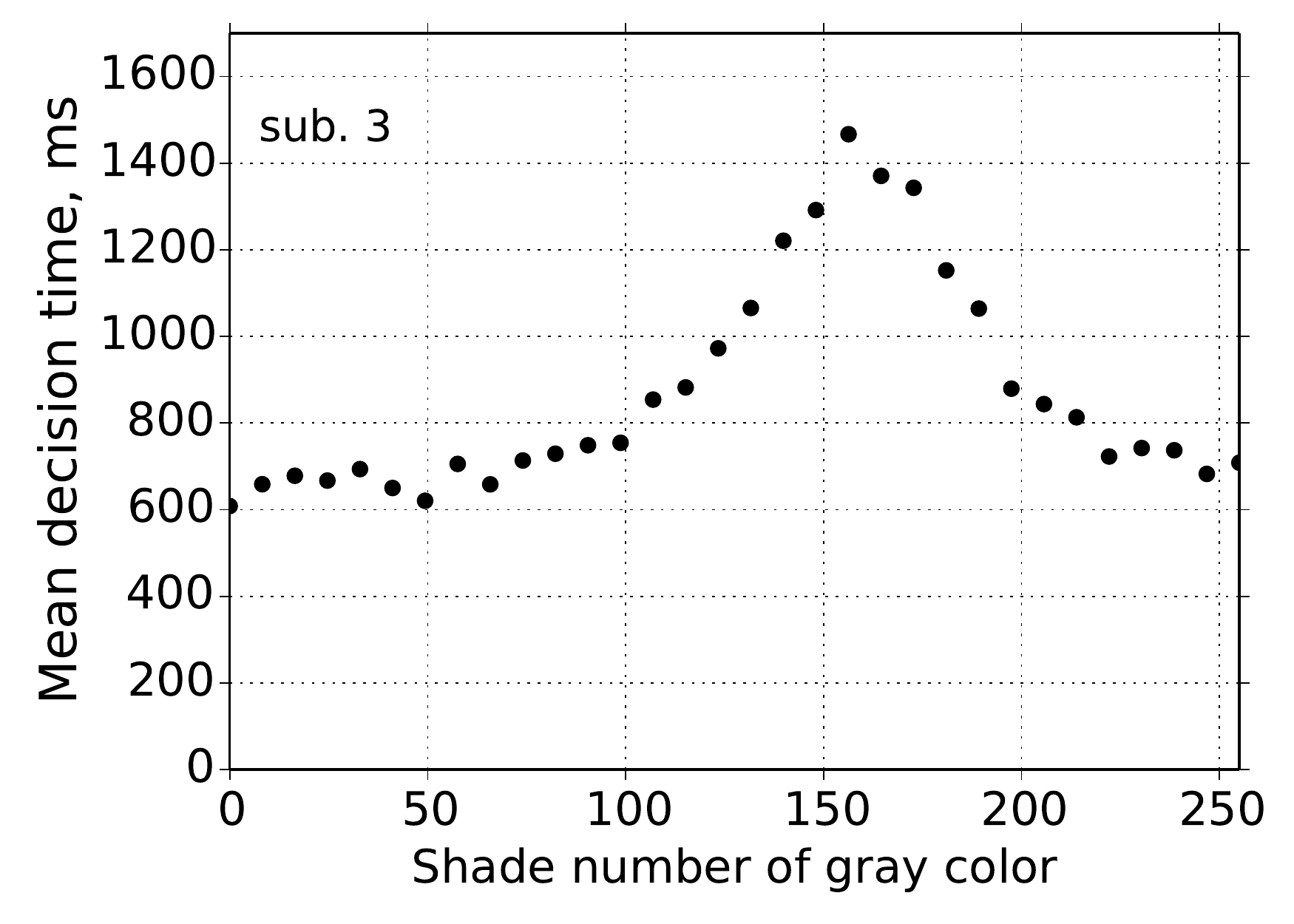}\quad\quad
		\includegraphics[width=\figscale\columnwidth]{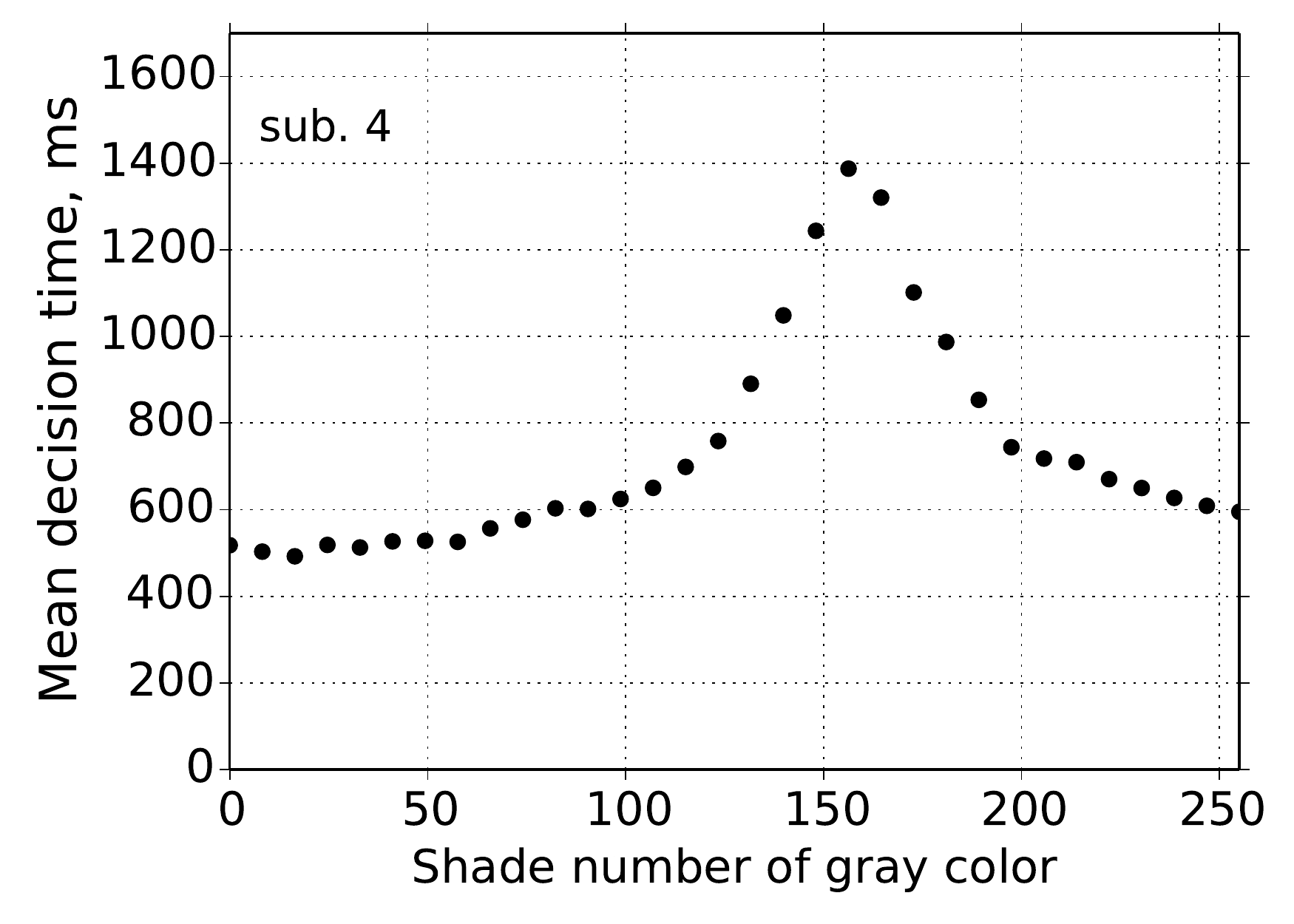}
	\end{center}
	\vspace*{-1.5\baselineskip}
	\caption{Mean decision time in gray color categorization depending on the shade number of gray.}
	\label{F4}
\end{figure*}

\section{Result and Discussion}

Figures~\ref{F3} and \ref{F4} present the obtained results. The psychometric functions of gray color categorization are depicted in Fig.~\ref{F3} for all the four subjects. The logarithmic scale of the $P$($y$)-axis is used to visualize the tails of their dependence on the shade number of gray color, $I$. As demonstrated in Fig.~\ref{F3}, the asymptotic behavior the constructed functions $P_{w,g}(I)$ can be approximated by 
\begin{equation}\label{fitfun}
P_{w,b}(I) \approx \frac12 \left\{1+ \tanh\left[\pm \frac{(I - I_{m:w,b})}{\delta I_{w,b}}\right]\right\},
\end{equation}
where $I_{m:w,b}$ is the center point of the crossover region and the value $\delta I_{w,b}$ characterizes its thickness. The used parameters of this approximation in fitting the experimental data are given in Table~\ref{Tab:1} for each subject individually.

As seen in Fig.~\ref{F3}, the asymptotics of both the psychometric functions looks like exponential function of the shade number, $P_{w,b}\propto \exp(\pm I/\delta I_{w,p})$, what is the typical dependence for the model of random systems residing in a bimodal potential well near the equiprobable  distribution. In this case the difference $\delta U$ in the well minima leads to the asymmetry in the well populations quantified by their ratio proportional to $\tanh(\delta U/T)$ (here $T$ is some constant). As follows from Tab.~\ref{Tab:1} the asymptotics of the psychometric functions at both the sides of the crossover region are characterized practically by the same parameters, $I_{m:w}\approx I_{m:b}$ and $\delta I_w\approx \delta I_b$ for subjects individually except for Subject~2. First, it allows us to suppose that the initial gamma correction of the monitor screen is acceptable at least in some neighborhood of the crossover region. The found asymmetry for Subject~2 can be explained by a relatively large width of the crossover such that the nonlinear properties of human brightness perception, in particular, growth of the brightness variation threshold with the brightness increase \cite{gescheider2013psychophysics},   give rise to its deeper ``penetration'' into the ``light-gray'' region. Second, the found results can be regarded as some argumentation for the hypothesis that categorical perception, at least, in the analyzed case is described by a certain potential model, where the corresponding decision-making is regarded as some random process $\eta$ in a potential relief $U(\eta|I)$ with two minima $U_w(I)$ and $U_b(I)$ determined by the shade number $I$ as a control parameter. In particular, in this case the probability $P_{w,b}$ of choosing LG and DG categories should be proportional to
\[
P_{w,b} \propto \exp\left\{ -{U_{w,b}(I)}\right\}.
\]
Verification of this hypothesis for the choice dynamics is an individual problem requiring, for example, an investigation of color categorization of shades of gray changing gradually in time.

\begin{table}
	\caption{Parameters used in fitting the data presented in Fig.~\ref{F3} by expression~(\ref{fitfun}).}
	\label{Tab:1}
	\begin{center}
		\begin{tabular}{c|cccc}
			subject ID & $I_{m:w}$ & $\delta I_w$ & $I_{m:b}$  & $\delta I_b$  \\ \hline
			1       &  181  &     16       &   182  &      18       \\
			2       &  138  &     22       &   140  &      29       \\
			3       &  157  &     22       &   155  &      23       \\
			4       &  156  &     15       &   155  &      14        
		\end{tabular}
	\end{center}
\end{table}

The experimental data shown in Fig.~\ref{F3} for all the subjects contain two domains on both the sides of the crossover region that may be classified as domains of scattered data that are caused by some mechanism rather than lack of statistics in the collected data. It is most pronounced for Subject~2 where these scattered data demonstrate evident growth in comparison with the crossover data. In order to clarify a plausible mechanism that could be responsible for this anomalous behavior of these heavy tails of the psychometric functions, we analyzed the dependence of the mean time required for subjects to make decision about classifying a current shade of gray depending on the shade number $I$. The results are exhibited in Fig.~\ref{F4}.  For all the subjects these distributions exhibit remarkable peak located inside the crossover region.
Outside this peak the values of the mean decision time $T$ are located in the interval 500--700~ms, which can be regarded as the upper boundary of the human reaction delay time controlled by physiological processes of recognizing threshold events within their unpredictable appearance. At least, it is the upper boundary of visual time intervals presenting timescales relevant to natural behavior, see, e.g., \cite{Mayo22012013} and references therein.  The mean decision time in these peaks is about 1.5~s (except for Subject~2 for which this time is 1~s). To explain these values we pose a hypothesis that mental processes on their own rather than pure physiological mechanisms contribute substantially to the decision making in color categorization under pronounced uncertainty, at least, in the analyzed case. As far as the scattered data domains in Fig.~\ref{F3} are concerned, we explain their appearance by addressing this phenomenon to a relatively short time of decision when the right choice is rather evident, which can increase the probability of subjects pressing a wrong button accidentally. A similar dependence of the decision time was found in the speech recognition \cite{bidelman2013tracing}, however, the time delay attained in the peak maximum does exceed 600~ms. In our data the mean time of decision-making includes also a time interval between making decision and pressing the corresponding button. However, if the time delay is deducted from the measured time data, the found peak will be even more pronounced. 
Summarizing this argumentation we pose a hypothesis that the characteristic time scale of decision-making in categorical perception depends substantially on the uncertainty of the current choice.             

\section{Shape categorization}\label{Kobayashi} 

\begin{figure*}[t]
	\begin{center}	
		\includegraphics[width=0.95\columnwidth]{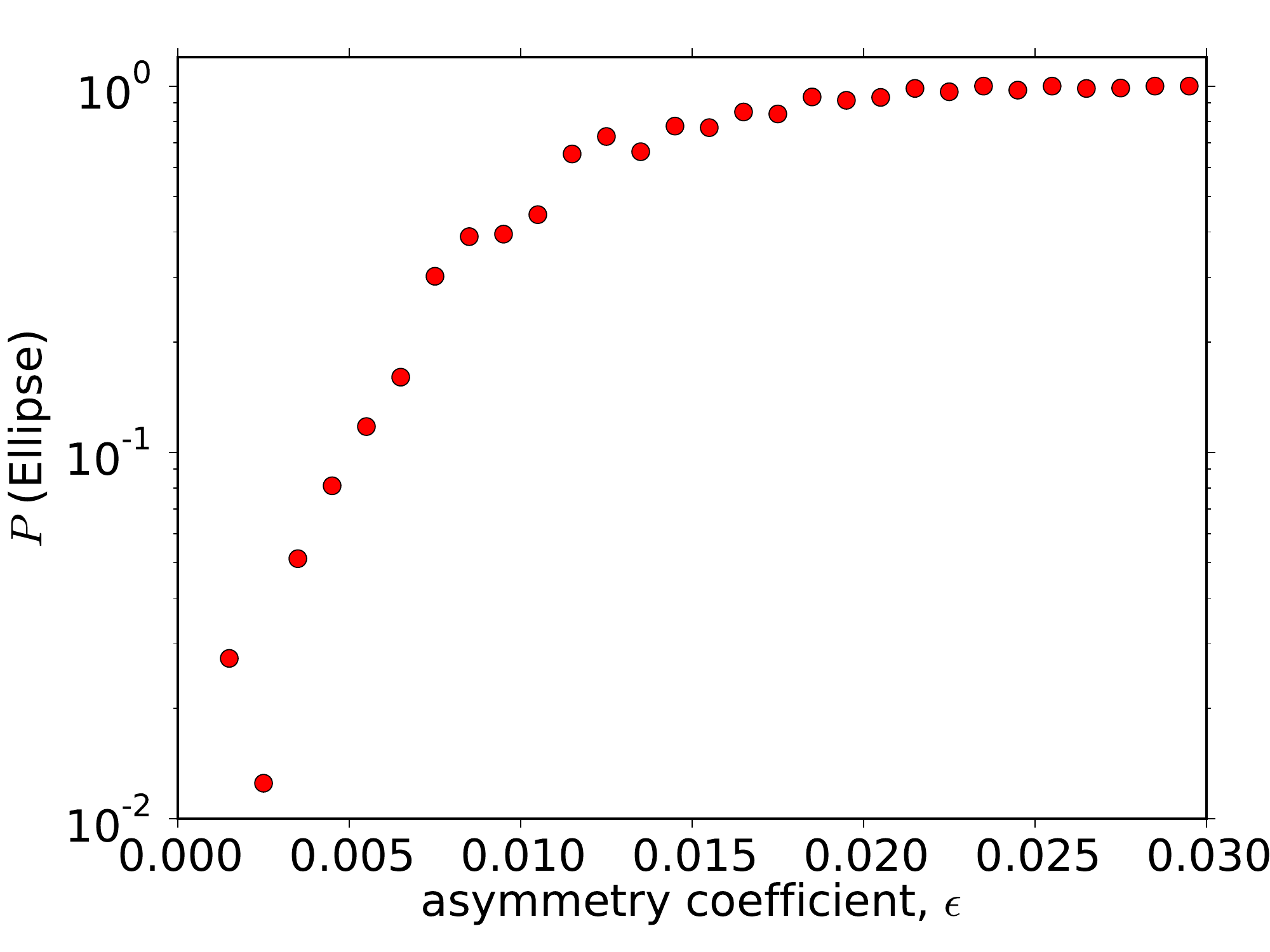}\quad
		\includegraphics[width=0.95\columnwidth]{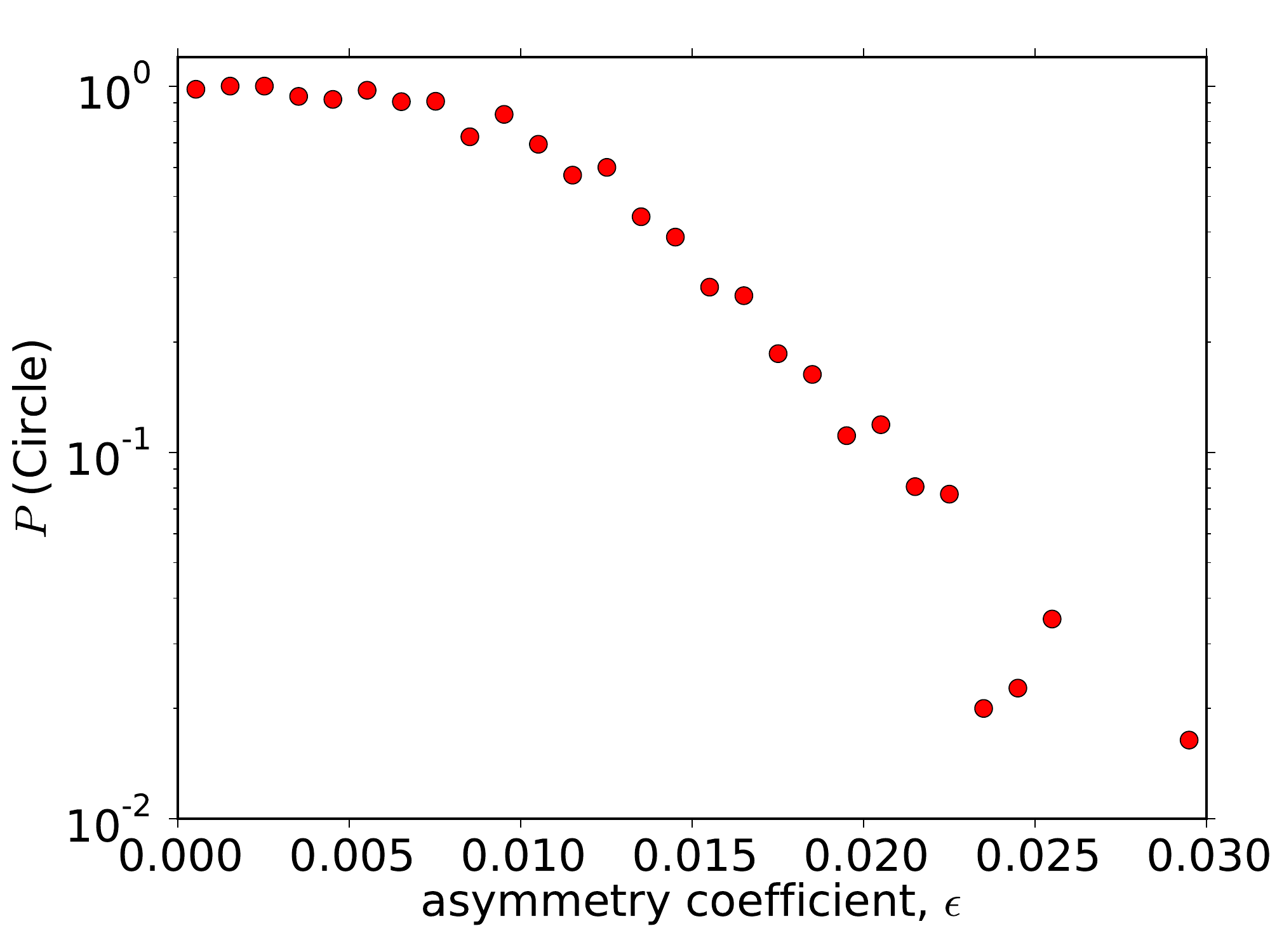}
	\end{center}
	\caption{Psychometric functions constructed for the collected data for one of the subjects. The frames present the probability of classifying the ellipse with asymmetry coefficient $\epsilon$ as asymmetric ellipse, $1 - P(\epsilon)$, or circle, $P(\epsilon)$, respectively. These plots have practically the same form for all the subjects.}
	\label{fig:data1}
	\begin{minipage}[b]{0.95\columnwidth}
		\begin{center}	
			\includegraphics[width=\columnwidth]{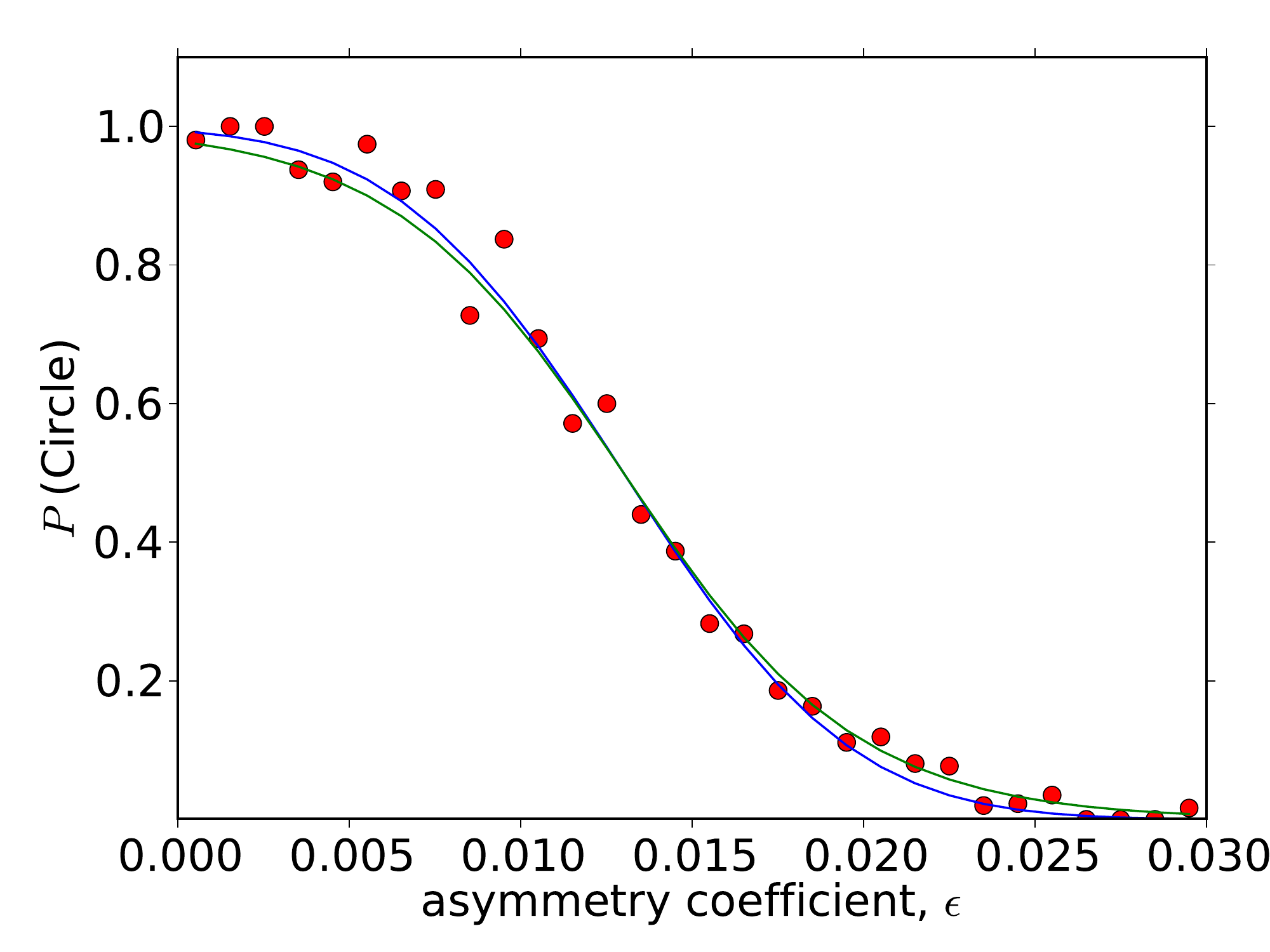}
		\end{center}
	\end{minipage}\quad\quad	
	\begin{minipage}[b]{1.07\columnwidth}
		\caption{Psychometric function of the ``circle'' categorization constructed based on the collected data in normal scales and the fitting curves. The blue line shows the approximation of the psychometric function $P(\epsilon)$ using the error function~\eqref{fit1} and the green line represents the logistic function approximation~\eqref{fit2}. The used parameters of these approximations are $\epsilon_\text{th,1} = \epsilon_\text{th,1} =  0.013$, $\ell_1= 0.0074$, and $\ell_2 =  0.0068$.}
		\label{fig:data2}	
		\vspace*{0.5\baselineskip}
	\end{minipage}
\end{figure*}

This supplementary section presents a short description of the results obtained in our pilot experiments of shape recognition \cite{Kobayashi2014}. In these experiments a sequence of ellipses with randomly generated asymmetry were visualized in a way similar to one described above. Each trial an ellipse whose axes are rotated by 45 degree is generated with the direction of the larger axis chosen randomly. Three subjects (male students  of 22 years old) involved in these experiments had to classify a visualized curve as ``circle'' or ``asymmetric ellipse.'' The subjects were instructed to note the direction of the larger axis in categorizing the ellipse asymmetry or classify it as ``circle,'' which models the situation when humans cannot distinguish between the strict circle and an ellipse with weak asymmetry.

The ellipse asymmetry is quantified by the variable $\epsilon$ related to the ellipse axes as 

\begin{align*}
	W & = R  \sqrt{\frac{ (1 - {\epsilon}) } { (1 + {\epsilon})}}\,, &
	H &= R  \sqrt{\frac{ (1 +{\epsilon}) } { (1 - {\epsilon})}}\,,
\end{align*}
where $R$ is a fixed quantity about 10~cm on the screen.

In order to accumulate the required amount of statistics and to depress uncontrollable fatigue effects the experiments were organized as the set of four trials with one trial per day. During each trial 500 data-points and totally 2000 data-points set were collected.  

Figure~\ref{fig:data1} exhibits the constructed psychometric functions. As seen in this figure the asymptotic behavior of the found psychometric functions is actually described by the exponential function $e^{A\epsilon}$ (here $A$ is a constant). 

Figure~\ref{fig:data2} shows the psychometric function of the ``circle'' categorization in the linear scales and the solid curves representing  the following fitting models
\begin{align}
	\label{fit1}
	P_\text{1}(\epsilon) & = \frac12\text{erfc\,}\left(\frac{\epsilon-\epsilon_\text{th,1}}{\ell_1}\right)\,,\\
	\label{fit2}
	P_\text{2}(\epsilon) & = \frac12 \left[1-\tanh\left(\frac{\epsilon-\epsilon_\text{th,2}}{\ell_2}\right)\right]\,,
\end{align}
The parameters of these approximations were chosen in such way that the standard deviation
\begin{equation*}
	\sqrt{\frac1{N} \sum_{i\in \text{data}} \left[\Delta P_i - \left<\Delta P_i\right> \right]^2}
\end{equation*}
take its minimum. Here we have used the notation
\begin{equation*}
	\Delta P_i =    
	P^\text{data point}_i - P^\text{approximation}(\epsilon_i)
\end{equation*}
and $N$ is the number of data points obtained for the psychometric function. 
In particular, it poses a question about the criteria for the choice of the fitting approximations. If a given approximation is based on minimizing the standard deviation then the two functions~\eqref{fit1} and \eqref{fit2} are practically equivalent. It is due to the experimental points located inside the transition region mainly contribute to the standard deviation. However their asymptotic behavior is considerably different and, from this point of view, approximation~\eqref{fit2} is the best among them. 

\section{Conclusion}
The present work reports the results obtained in our experiments on categorical perception of shades of gray. For this purpose a special color generator was created that generates a random sequence of shades of gray. Four subjects were involved in these experiments and instructed to categorize displayed shades of gray into two classes, light-gray and dark-gray. The pivot point of our research is the analysis of the asymptotic behavior of the psychometric functions. To accumulate the required amount of statistics, special efforts were made, including 5 days of experiment with 1 hour continuous trials per each day.

The particular results argue for the following hypotheses.

\begin{itemize}
	\item Categorical perception, at least, of shades of gray, is governed by a potential mechanism of decision-making treated as a random process in a potential relief. The previous analysis of shape recognition \cite{Kobayashi2014} also reveled the same type asymptotics of psychometric functions. Noting that the cognitional perception and shape recognition should be governed by different mechanisms, the found universality allows us to pose an assumption that human decision-making under uncertainty is implemented via a common emergent mechanism. In this mechanism the uncertainty measure plays the role of a certain parameter aggregating in itself particular physiological details.  
	
	\item The characteristic time scale of decision-making during categorical perception, at least in the analyzed case, depends substantially on the uncertainty in classifying a current event; the higher the uncertainty, the longer the decision time. The obtained data enable us to relate this effect with considerable contribution of mental processes to categorization.  In this feature categorization perception differs from recognition process near perception thresholds seemed to be governed by physiological mechanisms only.   
	
	\item As demonstrated in the experiments on shape recognition \cite{Kobayashi2014}, different models for human perception can give rise to practically the same form of the corresponding psychometric functions within the normal scales. Therefore, keeping also in mind that exactly the tails of psychometric functions bear the information about the feature of the human recognition mechanism, fitting the deep tails could be regarded as an essential requirement in constructing an appropriate approximation.      
\end{itemize}

\appendix
\section{Dynamic optimization of random number generation}\label{App:B}

\begin{figure*}
	\begin{center}
		\includegraphics[width=2\columnwidth]{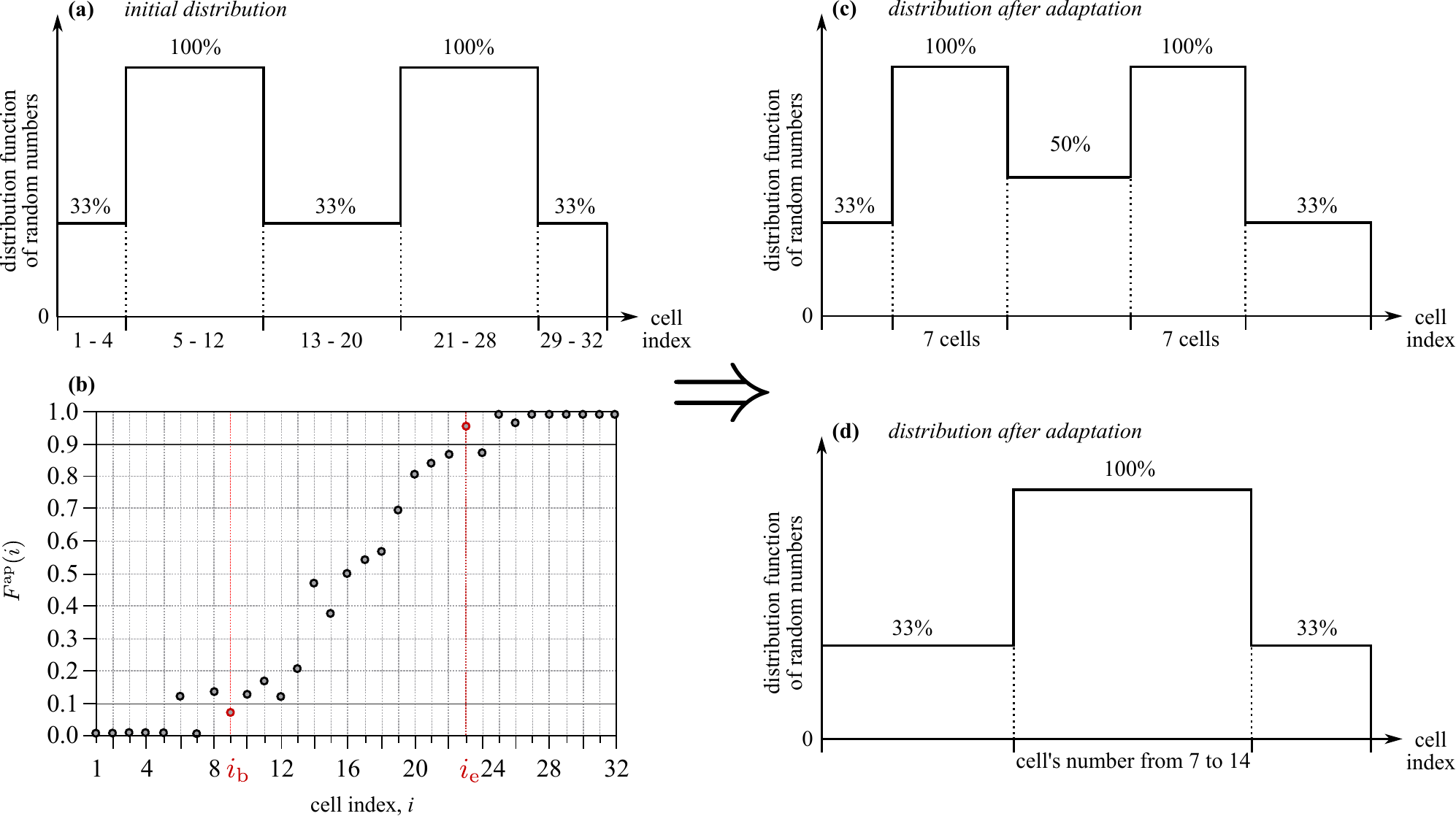}
	\end{center}
	\caption{Illustration of adapting the random number generator aimed at increasing the amount of experimental data inside the regions of main interest.}
	\label{Fig:RNGDF}
\end{figure*}

Based on preliminary conducted pilot experiments it was found that the psychometric functions to be constructed remain smooth enough with the sampling of 8 points per one cell; so the total interval of the possible values of the generated integers, [0,255], is divided into 32 cells of the same width. Then based on the class Random being the standard components of C\# distributive an adaptive generator with a bimodal distribution of random numbers was created. The main idea is that two maxima of this distribution function be located near the beginning and ending of the crossover region of the psychometric functions. At the beginning of each experiment the first 300 records are obtained using the random number generator with the distribution function shown in Fig.~\ref{Fig:RNGDF}(\textbf{a}); it matches the typical data collected during the preliminary experiments. These 300 records are used to construct a rough approximation $F^\mathrm{ap}(i)$ of one of the psychometric functions Fig.~\ref{Fig:RNGDF}(\textbf{b}); here $i$ is the cell index and the details of constructing the psychometric functions will be given in Sec.~\ref{sec:DA}.  Then the beginning of the crossover region is specified as the first cell $i_b$ preceding the \textit{first three successive cells} at which the constructed  psychometric function $F^\mathrm{ap}(i)$  exceeds 0.1 (10\% threshold) or deviates from unity by 10\%. The ending of the crossover region is specified in the same way except for the last three cells are determined and the following next one, $i_e$ is finally chosen. After that the random number generated is modified such that centers of its three peaks are placed at the beginning and ending of the crossover region. The final form of the random number distribution function is shown in Fig.~\ref{Fig:RNGDF}(\textbf{c,d}) depending on whether the peaks overlap each other or do not.



\end{document}